\documentclass{mn2e}
\usepackage{times,graphicx,amssymb,color}

\title[Planet-satellite pairs]{A dynamical study on the habitability of terrestrial exoplanets I: Tidally evolved planet-satellite
pairs}
\author[R. Brasser et al.]{R. Brasser$^1$, S. Ida$^2$ and E. Kokubo$^3$\\
$^1$ Institute for Astronomy and Astrophysics, Academia Sinica, Taipei 10617, Taiwan\\ 
$^2$ Department of Earth and Planetary Sciences, Tokyo Institute of Technology, $\bar{O}$okayama, Meguro district, Tokyo 152-8551,
Japan\\ 
$^3$ Division of Theoretical Astronomy, National Astronomical Observatory of Japan, Osawa district, Mitaka, Tokyo 181-8588, Japan\\ 
}
\begin{document}
\maketitle
\begin{abstract}
We investigate the obliquity and spin period of Earth-Moon like systems after 4.5~Gyr of tidal evolution with various satellite
masses ($m_s = 0.0025m_p$ -- $0.05m_p$ where $m_p$ is the planet mass) and initial planetary obliquity ($\varepsilon_0 = 0^\circ$ --
175$^\circ$) and discuss their relations to the habitability of the planet. The satellite initially orbits in the planet's equatorial
plane at $\sim4$ planetary radii and the planet's initial rotation period is 5~h. The other tidal parameters are modelled after the
Earth and Moon and we keep the satellite on a circular orbit. We find three possible outcomes: either i) the system is still evolving,
such as our own, ii) the system is in the double synchronous state, with the planet's obliquity at either 0$^\circ$ or 180$^\circ$, or
iii) the satellite has collided with the planet. The case iii) occurs for initial planetary spins in the range $\varepsilon_0 \sim
60^\circ$ -- 120$^\circ$. For other $\varepsilon_0$, the satellite survives. The transition between case i) and ii) is abrupt and
occurs at slightly larger satellite mass ($m_s \sim 0.02m_p$) than the lunar mass. For higher masses the system is in the double
synchronous state and the final planetary spin periods ($P_p$) are longer than 96~h. We also discuss the habitability of the planet in
each case. We suggest that cases ii) and iii) are less habitable than case i). Using results from models of giant impacts and satellite
accretion, we found that the systems that mimic our own i.e., with rotation period $12 < P_p < 48$~h and current planetary obliquity
$\varepsilon_p < 40^\circ$ or $\varepsilon_p > 140^\circ$ only represent 14\% of the possible outcomes. This esimate may only
be reliable to within factors of a few, depending on how the probability is evaluated. Elser et al. (2011) conclude that the
probability of a terrestrial planet having a heavy satellite is 13\%. Combining these results suggests that the probability of
ending up with a system such as our own is of the order of 2\%.
\end{abstract}

\section{Introduction}
With the recent detections of possible terrestrial planets in the habitable zones of other stars by the Kepler mission (Lissauer et
al., 2012) a question arises: to what extent could these remote worlds support life similar to what we know here on Earth
(terrestrial-type life)? There are many conditions for a planet to fulfil to support terrestrial-type life, which can be controlled by
either geological, climatological, orbital, or geophysical processes. In this study, the first in a series, we focus on the dynamical
factors that may affect the habitability of a terrestrial exoplanet. We identify several key features that we believe are
important, if not essential, for supporting terrestrial-type life. These are i) a stable climate, ii) low to moderate ($<$
50$^\circ$) seasonal temperature variations, iii) low to moderate diurnal temperature variations, and iv) low to moderate spatial
variation of temperature over the planet. These conditions need to be supplemented with the following: v) regular, low-amplitude
obliquity oscillations, vi) moderate obliquity to induce seasonal variations, vii) moderate rotation rate, and viii) small orbital
perturbations. We examine each of these below.\\

For the current Earth, the first condition requires a constant obliquity and a low eccentricity. Earth's life is both land-based
and water-based and requires a stable climate over millions or even billions of years. In addition to continental drift the long-term
climate on the Earth is largely regulated by the Milankovi\'{c} cycles (Milankovi\'{c}, 1941). These cycles are related to variations
in Earth's orbit and their influence on the obliquity. The main cycles are related to its equatorial precession (with period 26~kyr,
corresponding to a frequency $\dot{\psi}=$-50.5~''/yr, where $\psi$ is the angle between the vernal equinox and the intersection of the
equator with the ecliptic), the obliquity variations of the Earth (with period 41~kyr, corresponding to the frequency $\dot{\psi}-s_3$,
where $s_3$ is the Earth's nodal eigenfrequency), and the orbital eccentricity (with periods 100~kyr and 400~kyr, corresponding to
frequencies $g_2+g_5$ and $g_2-g_5$, where $g_2$ and $g_5$ are the eccentricity eigenfrequencies of Venus and Jupiter respectively).
The present obliquity variations of the Earth are small: the amplitude is just 1.5$^\circ$, and the low amplitude is a result of the
presence of the Moon. However, even these small oscillations are enough to cause regular ice ages through the positive feedback effect
(Deser et al., 2000).\\

The second condition that is necessary to make the planet desirable for terrestrial-type life leans on the following. Spiegel et
al. (2009) have shown that the polar regions of a planet with perpendicular spin are mostly uninhabitable because of the lack of
seasons: the cold regions will mostly stay cold, possibly even below freezing. On Earth the onset of the ice ages tends to favour low
obliquity for two reasons: the reduction in overall summer insolation at high latitudes and the corresponding reduction in mean
insolation (Huybers \& Wunsch, 2005). The cooling would result in more ice building up near the polar caps, increasing the albedo which
then instigates a positive feedback effect. However, there are other theories that show that precession and eccentricity forcing on the
insolation also play a role (e.g. Imbrie \& Imbrie, 1980; Paillard, 1997; Huybers, 2011) and the debate is ongoing. The weak seasonal
effects at low obliquity and the corresponding constant low temperatures at high latitudes may imply that habitability is increased for
planets that have a moderate obliquity because the associated seasons make the polar regions partly habitable by increasing the yearly
mean insolation, and thus the temperature. However, too high an obliquity may not be favourable for habitability either (Williams \&
Kasting 1997; Williams \& Pollard, 2003; Spiegel et al., 2009) because the long summers at the poles could cause large seasonal
temperature variations above large continents (Williams \& Pollard, 2003) and the highest temperatures could exceed 50~$^\circ$~C.
Additionally the long periods of darkness even at mid latitudes could cause difficulties for photosynthetic life to survive. Thus it
appears that the current obliquity favours the development of terrestrial-type life rather than more extreme values.\\

The third condition, a low to moderate diurnal temperature variation, is regulated by a sufficiently slow rotation rate, a thick
atmosphere and oceans. Diurnal temperature changes are partially governed by the relaxation time scale for atmospheric heat losses and
temperature variations. On the Earth this time scale is 20 days (Matsuda, 2000) and the diurnal temperature variation is approximately
10~K. For Mars the story is very different: the relaxation time is comparable to its rotation period and the diurnal temperature
variation is approximately 60~K (Matsuda, 2000). Part of the reason the diurnal variation on Mars is much higher than on Earth is
because the latter's oceans store heat much more effectively than land and Mars' atmosphere is much more tenuous. However, a
very slow rotation will cause the Hadley cell to cover the whole planet (Farrell, 1990) and the diurnal temperature variations may be
diminished. A fast rotation, on the other hand, would decrease the heat transport as argued in the next paragraph. By themselves
these temperature changes may not pose a problem but, coupled with the seasonal variations, they may be difficult for land-based life
to adapt to. \\

The fourth condition, low to moderate spatial variation of temperature over the planet, can be the result of the following. 
Simulations of global circulation models on an Earth-like planet with different rotation rates have shown that for a fast rotating
Earth the atmosphere experiences the creation of many small Hadley-like cells (Williams, 1988a) and most of the heat transport is
caused by baroclinic eddies. These eddies decrease the efficiency of heat transportation from the warmer regions to the colder regions.
This possible reduction in heat transportation by the atmosphere may decrease the habitability of the planet because the polar (and
thus colder) regions would probably stay cold (Williams, 1988a). However, even though there appears evidence for a more equable climate
having existed during the Cretaceous and Eocene aeons (e.g. Barron, 1989), which may have been caused by latitudinal ocean currents
rather than the present-day longitudinal ones (Bice \& Marotzke, 2002), it is not clear to what extent the size of the Hadley cells
depends on rotation rate or on land mass distribution, and how this increased equator to poleward heat flow could have been
sustained (Barron, 1983). Nevertheless it is probable the area of habitability is smaller on a fast-rotating planet than it is on a
slow-spinning one. Secondly, we have the issue that pertains to the effect of tides on ocean mixing. It is well known that tidal
forcing causes the mixing of stratified layers in the ocean (e.g. Egbert \& Ray, 2000), though recent work shows that mixing by
swimming sea creatures could be equally important (Katija \& Dabiri, 2009). Whatever its source this mixing has profound effects on the
Earth's climate because it allows for more efficient energy exchange between the atmosphere and the ocean, when cooler water reaches
the surface from the deeper regions. The mixing also aids in the transport of water from warmer regions to colder ones by currents such
as the Gulfstream (Garrett, 2003). Last, the mixing brings up important nutrients from the depths of the ocean that micro-organisms
residing closer to the surface can feed on. Thus, it appears that in a simplistic sense ocean mixing could be an important ingredient
in the sustenance of terrestrial-type life, and the habitability of the Earth might be decreased without this effect.\\

Regarding the second set of conditions for the existence of terrestrial-type life, v), vi) and vii) are satisfied because of the
presence of the Moon, while conditions v) and viii) are the result of the planets all having small eccentricities and mutual
inclinations, and the Earth being far from a secular orbital resonance and from a (secular) spin-orbit resonance. Thus at first glance
it appears that the Moon plays a key role in supporting life on Earth. However, some of the above definitions should be
interpreted with care. Mars appears to fulfil several of these criteria (moderate rotation speed, moderate obliquity, moderate orbital
perturbations, moderate seasonal temperature variations). In addition, it could have a stable spin-axis if the secular frequencies
where different. However, the slow rotation comes from the fact that it is a planetary embryo (Dauphas \& Pourmand, 2011), and embryos
appear to be less adapted for life because they lack the gravitational strength to keep a thick atmosphere and oceans against Jeans
escape, impact erosion or stellar wind pick-up over 4.5 Gyrs. The mass of an embryo may be regulated by an `isolation mass', which is
one order smaller than an Earth mass at 1~AU. Thus for the remainder of this paper we focus on a fully-formed planet with a satellite,
such as the Earth-Moon system. But is the Earth-Moon system unique in its architecture or is the current system a common outcome? We
aim to answer that question and relate the findings to the habitability of the resulting systems. \\

It is generally believed that the Moon formed through the giant impact of a Mars-sized body with the proto-Earth (e.g. Hartmann \&
Davis, 1975; Cameron \& Ward, 1976; Ida et al, 1997; Kokubo et al., 2000; Canup, 2004). This impact melted the impactor and part of the
proto-Earth, and resulted in much of the ejecta orbiting the Earth (Benz et al., 1991; Cameron, 1997; Canup, 2004). The Moon
subsequently accreted from this disc of debris in less than a year (Ida et al., 1997). Usually the final disc mass was less than 2.5
lunar masses and orbited inside the Earth's Roche limit, currently at 2.9 Earth radii $(R_{\oplus})$. The Moon ends up orbiting in the
Earth's equatorial plane. Tidal evolution over the next 4.5~Gyr pushed it towards its current orbit (Goldreich, 1966; Touma \&
Wisdom, 1994). With the aid of numerical simulations Elser et al. (2011) conclude that binary planetary systems, such as the
Earth-Moon system, occur approximately 8\% of the time, ranging from 2.5\% up to 25\%. However, their studies did not take subsequent
tidal evolution into account.\\

Goldreich (1966) and Touma \& Wisdom (1994) have shown that the Earth's current obliquity of 23.5$^\circ$ requires its obliquity to
have been approximately 10$^\circ$ just after impact. Similarly, Earth's rotation period was approximately 5~h after the
impact and the Moon's inclination with respect to the Earth's equator was about 10$^\circ$. The initial spin of the Earth being nearly
perpendicular to its orbital plane is a fairly rare outcome after a Moon-forming impact: it has been shown that the obliquities of the
terrestrial planets are isotropically distributed (Chambers, 2001; Kokubo \& Ida, 2007) and their resulting rotation rates are
approximately half of their maximum values (Kokubo \& Genda, 2010). It is likely that the planetary rotation period is partially
determined by the impact that formed the satellite (Lissauer, 2000) and the satellite mass is related to the impact parameter of the
collision: grazing impacts produce heavier satellites and speed up the planet's rotation rate much more than head-on collisions (Kokubo
et al., 2000). This result suggests there is a relation between the planet's rotation period and the satellite that has formed.
However, as we show in Appendix A, the relation is not one-on-one but rather a rough estimate. Additionally, an isotropic obliquity
distribution favours coplanar spins rather than perpendicular ones. What is then the evolution of the Earth-Moon system if the Earth
had been much more oblique just after the Moon-forming impact?\\

\begin{figure}
\resizebox{\hsize}{!}{\includegraphics[]{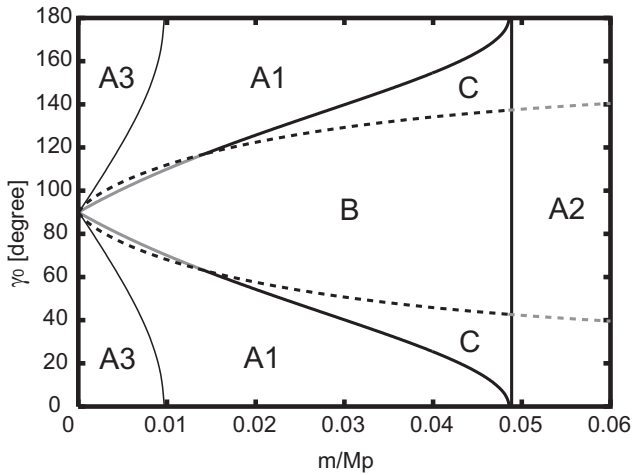}}
\caption{This figure shows the regions of the three possible outcomes of planetary motion as a function of satellite mass and original
planetary obliquity. Taken from Atobe \& Ida (2007). See text for details.}
\label{atobeida}
\end{figure}

Atobe \& Ida (2007) studied the tidal evolution of systems consisting of an Earth-like terrestrial planet with a satellite ranging from
sub-lunar to super-lunar mass on a circular orbit. They varied the initial mass of the satellite ($m_s$) and the initial obliquity of
the planet ($\varepsilon_0$) but kept the initial semi-major axis of the satellite at 3.8~$R_{\oplus}$, the Earth's rotation period at
5~h and the satellite's inclination with respect to the Earth's orbit, $i$, at 1$^\circ$. They showed that there are essentially three
outcomes when these systems are evolved towards completion: i) the satellite recedes from the planet and approaches (and sometimes
reaches) the outer co-rotation radius (case A), ii) the satellite first recedes from and then approaches the planet resulting in
collision (case B), or iii) the satellite first recedes from the planet and then approaches it but gets caught at the inner co-rotation
radius (case C). In general terms, case A occurs for low-mass satellites at low prograde or high retrograde obliquities. Case B occurs
for initial obliquities $\varepsilon_0 \in (\sim 60^\circ, \sim 120^\circ)$ and case C occurs for heavy satellites ($m_s\gtrsim
0.03\,m_p$, with $m_p$ being the planetary mass) at low obliquity. We have summarised the results in Fig.~\ref{atobeida}, taken from
Atobe \& Ida (2007), where the symbol $\gamma_0$ is $\varepsilon_0$ in our notation. The outcome case A is further subdivided
according to whether the system is still evolving. For case A1 prograde planetary spins will have the planetary obliquity,
$\varepsilon_p$ evolve to $\varepsilon_p \rightarrow 0^\circ$ while retrograde spins will evolve to $\varepsilon_p \rightarrow
180^\circ$. In the outcome A2 the system has stopped evolving. Outcome A3 is a particular case of A1, where a retrograde-spinning
planet will eventually spin prograde due to the dominance of solar tides over satellite tides (for A1 this does not happen). \\

From this figure it appears that the outcome for oblique planets can be very different from the current Earth-Moon system. Atobe \& Ida
(2007) studied the evolution of these systems until their ultimate state. For light satellites, such as the Moon, the time scale to
reach the final configuration is much longer than the age of the Solar System, exceeding the Hubble time for masses below half a lunar
mass. Thus their results cannot be used to directly determine the state of the system at the current epoch i.e. 4.5~Gyr after its
formation.\\

In this paper we determine the state of fictitious systems consisting of an Earth-like terrestrial exoplanet and a satellite whose
mass ranges from sub-lunar to super-lunar values. We investigate the tidal evolution of these systems starting with different initial
planetary obliquities and satellite masses. We do this by performing a series of simulations similar to those done by Atobe \& Ida
(2007) but we cease them after the system has reached an age of 4.5~Gyr. We then analyse the states of each system and determine which
regions of the phase space are the most suitable for supporting terrestrial-type life. In addition we attempt to determine how likely
it is to produce a system similar to our own from a range of initial conditions. This paper is structured as follows.\\

In the next section we lay out the equations of motion consisting of the star, planet and satellite. These are based on the work of
Bou\'{e} \& Laskar (2006) and Correia (2009). In section 3 we describe our numerical methods and compare the evolution from our
simulations to earlier results presented in Bou\'{e} \& Laskar (2006) for the conservative motion, and Touma \& Wisdom (1994) for the
tidal evolution. In section 4 we present the results of our numerical simulations, focusing on the final states of the system and how
these may change with slightly different initial conditions and passages through secular spin-orbit resonances. In the last section we
present a summary and conclusions.

\section{Theory of precession and tidal evolution}
We study the evolution of a hierarchical system consisting of a central star with mass $M_*$, a terrestrial planet with mass
$m_p$ and a satellite with mass $m_s$. The planet and satellite are oblate spheroids with zonal harmonics $J_2 = (C-A)/m_pR^2$, where
$R$ is the equatorial radius of the body under consideration and $C$ and $A$ are the principal and secondary moments of inertia. The
planet and satellite rotate about the axis with maximal inertia, $C$. From now on, we shall use the dimensionless variant
$\mathcal{C}=C/m_pR^2$. We follow Bou\'{e} \& Laskar (2006) and Correia (2009) in the derivation below.\\

The conservative motion of the system can be tracked through the conservation of the total angular momentum

\begin{equation}
{\bf \dot{L}_p + \dot{L}_s +\dot{H}_p + \dot{H}_s} = 0,
\label{angmomcons}
\end{equation}
where a dot denotes a time derivative, ${\bf H}_i = \mathcal{C}_im_iR_i^2\nu_i{\bf s}_i$ are the spin angular momenta of the planet and
satellite and ${\bf L}_i=m_in_ia_i^2(1-e_i)^{1/2}{\bf k}_i$ are their orbital angular momenta. Here $\nu_i$ is the spin frequency of
either body, $a_i$ are the semi-major axes, $e_i$ are their eccentricities and $n_i$ are their mean motion. The vectors ${\bf s}_i$ are
the unit vectors in the direction of the spin angular momentum of the planet and the satellite while the vectors ${\bf k}_i$ are the
unit vectors of the orbital angular momenta. In what follows the subscript $i$ can refer to either the planet or the satellite. Tidal
evolution acts on time scales much longer than the orbital period of both the planet and the satellite, so equation (\ref{angmomcons})
is averaged over the mean anomalies of both bodies and the periapse of the planet-satellite pair. The resulting equations for the
conservative motion are then given by (Bou\'{e} \& Laskar, 2006)

\begin{eqnarray}
\label{cons}
 {\bf \dot{H}}_i & = & -\alpha_i \cos \varepsilon_i \, {\bf k_p \times s_i} -\beta_i \cos \theta_i \, {\bf k_s \times s_i}, \nonumber
\\
{\bf \dot{L}}_s & =& -\gamma \cos I \, {\bf k_p \times k_s} + \sum_i \beta_i \cos \theta_i \, {\bf k_s \times s_i}, \\
{\bf \dot{L}}_p & =& \gamma \cos I \, {\bf k_p \times k_s} + \sum_i \alpha_i \cos \varepsilon_i \, {\bf k_p \times s_i}. \nonumber
\end{eqnarray}
The equations above assume that the eccentricity of the satellite is constant during one full rotation of the line of apses. This does
not necessarily have to be true, but we shall assume that it is so in this simplified model. A study that also accounts for the
eccentricity evolution of the satellite is left for the future. \\

In equations (\ref{cons}) above we have introduced the angles $I = \arccos({\bf k_p \cdot k_s})$, which is the inclination of the orbit
of the satellite with respect to the planet's orbital plane, $\varepsilon_i = \arccos({\bf k_p \cdot s_i})$ are the obliquities of both
bodies with respect to the planet's orbital plane and $\theta_i = \arccos({\bf k_s \cdot s_i})$ are the obliquities of both bodies with
respect to the satellite's orbital plane. In this study we are primarily interested in the quantities $I$, $\varepsilon_p$ and
partially $\theta_s$. We also introduced the precession constants

\begin{eqnarray}
\alpha_i &=& \frac{3GM_*m_iJ_{2i}R_i^2}{2a_p^3(1-e_p^2)^{3/2}}, \nonumber \\
\beta_i &=& \frac{3Gm_pm_sJ_{2i}R_i^2}{2a_s^3(1-e_s^2)^{3/2}}, \\
\gamma &=& \frac{3GM_*m_sa_s^2(2+3e_s^2)}{8a_p^3(1-e_p^2)^{3/2}}. \nonumber
\end{eqnarray}
Roughly, the quantities $\alpha_i$ are the precession rates of the figures of the planet and satellite caused by the torques from the
Sun, the quantities $\beta_i$ are the precession rates of the poles of both bodies caused by mutual torques between them, and $\gamma$
is the precession rate of the nodes of the satellite's orbit on the planet's orbit.\\

On long time scales the system is subject to tidal forces, most notably between the planet and the satellite. In this study we adopt
the constant time delay model of Mignard (1979, 1980) for its simplicity, keeping the eccentricity of the planet and satellite
constant (but not necessarily zero). The tidal forces from the planet and the satellite on each other act to reduce their rotation
rates. The reduction in spin angular momentum is compensated by an increase in the orbital angular momentum of the satellite. The
averaged equations of motion governing the tidal evolution of the planet-satellite pair are (Correia, 2009)

\begin{eqnarray}
{\bf \dot{H}}_i &=& -{\textstyle{\frac{1}{2}}}K_i[X^{-6,0}_0(e_s)({\bf s}_i + \cos \theta_i {\bf k}_s)\nu_i  \nonumber \\
&-& 2n_sX^{-8,0}_0(e_s)(1-e_s^2)^{1/2}{\bf k}_s], \\
{\bf \dot{L}}_s &=& {\textstyle{\frac{1}{2}}}\sum_iK_i[X^{-6,0}_0(e_s)({\bf s}_i + \cos \theta_i {\bf k}_s)\nu_i  \nonumber \\
&-& 2n_sX^{-8,0}_0(e_s)(1-e_s^2)^{1/2}{\bf k}_s], \nonumber
\end{eqnarray}
where we introduced

\begin{eqnarray}
K_p = 3k_{2p}Gm_s^2R_p^5\Delta t_p a_s^{-6}, \nonumber \\
K_s = 3k_{2s}Gm_p^2R_s^5\Delta t_s a_s^{-6}.
\end{eqnarray}
Here $k_{2i}$ is the secular Love number of the planet or satellite, $\Delta t_i$ is the tidal delay and $X^{-6,0}_0(e)$ and
$X^{-8,0}_0(e)$ are Hansen coefficients. These are defined from

\begin{equation}
\Bigl(\frac{r}{a}\Bigr)^n\exp({\rm i}mv) = \sum_{l=-\infty}^{\infty}X_l^{n,m}(e)\exp({\rm i}lM),
\end{equation}
where $r$ is the distance of the planet to the star, $v$ is the true anomaly, $M$ is the mean anomaly, $l$, $m$ and $n$ are integers
and ${\rm i}^2=-1$ is the imaginary unit. The time delay is related to the tidal dissipation parameter, $Q$, via $\Delta t =
(Q\nu)^{-1}$.\\

The planet-satellite pair is not isolated and it will be subjected to tidal forces from the central star. For simplicity we only
incorporated the tidal action from the star on the planet, and thus the reduction in the planet's spin angular momentum is
compensated by an increase in its orbital angular momentum, giving

\begin{eqnarray}
 {\bf \dot{H}}_p &=& -{\textstyle{\frac{1}{2}}}N[X^{-6,0}_0(e_p)({\bf s}_p + \cos \varepsilon_i {\bf k}_p)\nu_p \nonumber \\
&-& 2n_pX^{-8,0}_0(e_p)(1-e_p^2)^{1/2}{\bf k}_p],
\end{eqnarray}
with $N=3k_{2p}GM_*^2R_p^5\Delta t_p a_p^{-6}$. As the satellite recedes from the planet and the rotation of both bodies slows
down, their $J_2$ coefficients decrease because the rotational deformation of their figures becomes less severe. The $J_2$ values of
the planet and satellite are updated according to $J_{2i} = \frac{1}{3}k_{si}R_i^3\nu_i^2/Gm_i$ (Atobe \& Ida, 2007), where $k_s$ is
the secular Love number, which is assumed to be constant. \\

Now that we have discussed the basic theory of the tidal and conservative motion, we describe our numerical methods below and
compare them with previous results.

\section{Numerical methods and tests}
In order to determine the evolution of the planet-satellite system with various initial conditions, we integrated the equations of
motion with the aid of computer codes. We wrote three versions. The first only integrates the conservative motion without any tidal
evolution. A second version only includes planet-satellite tides while the third version also includes the solar tides. The
integration was performed using the Bulirsch-Stoer method (Bulirsch \& Stoer, 1966). We checked our code against the results of Touma
\& Wisdom (1994) and Bou\'{e} \& Laskar (2006), which were all reproduced. \\

The input parameters are the masses, obliquities, semi-major axes, eccentricities, radii, moments of inertia $\mathcal{C}$, tidal
dissipation factors $Q$, Love numbers $k_2$, mutual inclination $i$, and rotation periods. The values of $J_2$ were derived from the
input values. The unit vectors were computed from

\begin{eqnarray}
\bf{k}_p &=& (0,0,1)^T, \\
\bf{k}_s &=& (\sin I \sin \Omega, -\sin I \cos \Omega, \cos I)^T, \\
\bf{s}_s &=& (\sin \varepsilon_p \sin \psi, -\sin \varepsilon_p \cos \psi, \cos \varepsilon_p)^T,
\end{eqnarray}
with $\bf{s}_s$ generated similar to $\bf{s}_p$. For a close satellite we set $\Omega = \psi$ and $I \sim \varepsilon_p$ so that $i
\sim 0$. In Table~\ref{tab1} we have listed all the input parameters and their initial values.\\

It is well known that integrating the tidal equations backward in time using the current tidal dissipation rate causes the
Moon to fall onto the Earth approximately 1~Gyr ago (Touma \& Wisdom, 1994), so we scaled the final integration time to be 4.5~Gyr to
obtain the current system state. In most cases we also lowered $Q_p$ to 10 to hasten the evolution and save CPU time. We made sure that
this did not affect the final outcome. \\

\begin{figure*}
\resizebox{\hsize}{!}{\includegraphics[angle=-90]{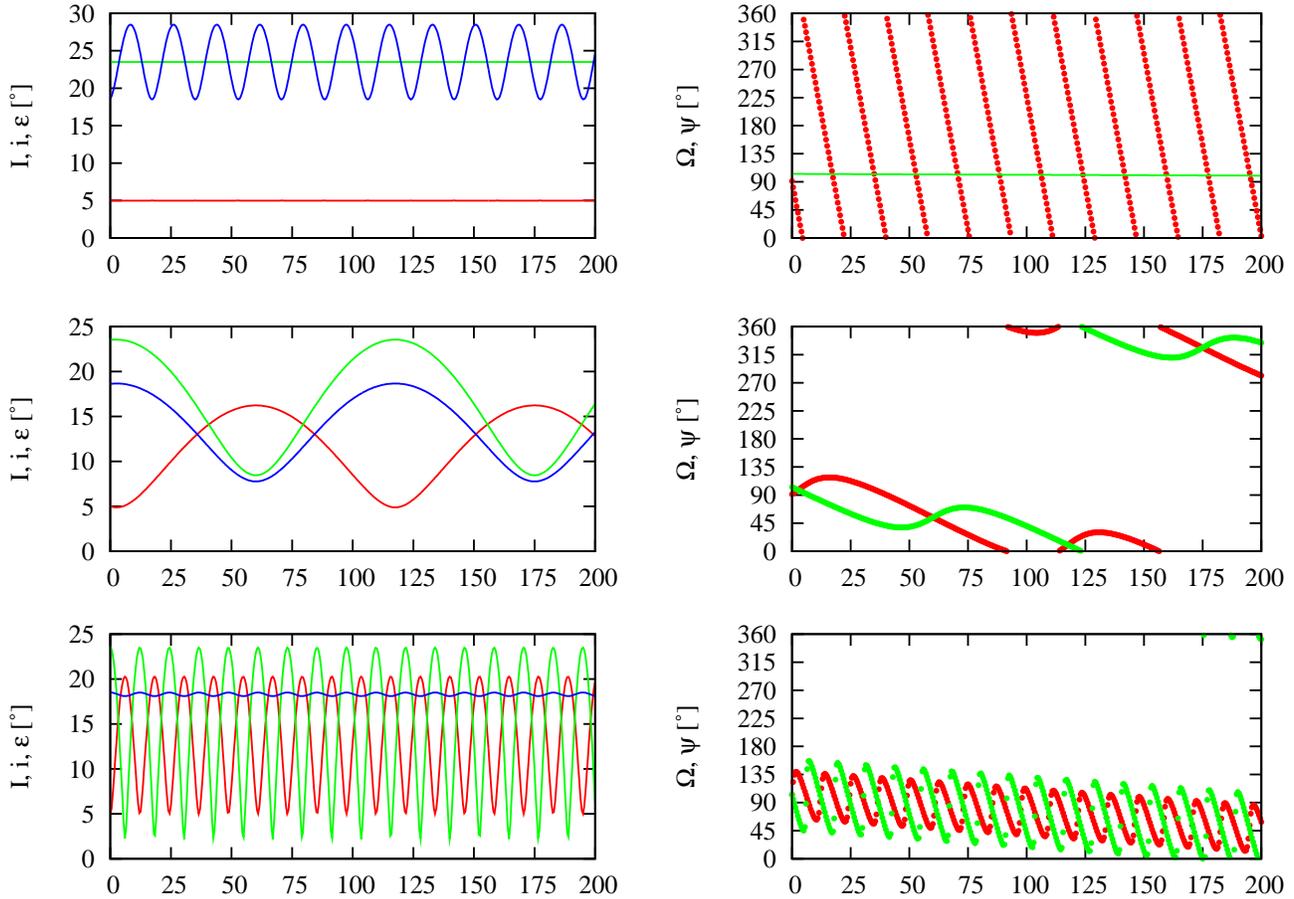}}
\caption{Orbital evolution of the Earth-Moon system for several different initial configurations. The top panels depict the current
state with the Moon at 60~$R_{\oplus}$. The middle panels pertain to the Moon being at 10~$R_{\oplus}$ and the bottom panels have the
Moon at 5~$R_{\oplus}$. On the left column we plot $I$ (red), $\varepsilon_p$ (green) and $i$ (blue). The right panels plot $\Omega$
(red) and $\psi$ (green).}
\label{em}
\end{figure*}

\begin{table}
 \begin{tabular}{ccc}
  Quantity & Planet & Satellite \\ \hline \\
  $m_i$ & $3.005 \times 10^{-6}$~$M_\odot$ & 0.0025 - 0.05 $m_p$ \\
  $R_i$ & 4.2634965 $\times 10^{-5}$~AU & 1.161848 $\times 10^{-5}$~AU\\
  $P_i$ & 5~h (3~h and 7~h) & 18.7~h \\
  $\varepsilon_i$ & 0$^\circ$-175$^\circ$ & variable\\
  $\theta_i$ & variable & 0 \\
  $\mathcal{C}$ & 0.331 & 0.394\\
  $k_2$ & 0.300 & 0.02264 \\
  $k_s$ & 0.95 & 1.2 \\
  $Q_i$ & 20 & 30 \\
  $a_i$ & 1.0~AU (0.75~AU and 1.25~AU) & 3.8~$R_\oplus$\\
  $e_i$ & 0 & 0\\
  $i$ & - & 0, 1$^\circ$ and 5$^\circ$
 \end{tabular}
\caption{Initial values of various quantities for our numerical simulations.}
\label{tab1}
\end{table}
To demonstrate the validity of the code and to show the different types of precession without tides, we plot the evolution of a
fictitious Earth-Moon system for various lunar distances in Fig.~\ref{em}. We have taken the current system and simply decreased the
lunar semi-major axis, keeping everything else the same. Thus the evolution at short lunar semi-major axis is unlikely to be
representative of the past lunar orbit but serves to illustrate the various types of motion of the system. \\

The top two panels are for the current system with $a_s = 60$~$R_\oplus$, the middle panels have $a_s = 10$~$R_\oplus$ and the bottom
panels are for $a_s = 5$~$R_\oplus$. The left panels plot the inclination of the lunar orbit with respect to the ecliptic ($I$) in red,
the inclination of the lunar orbit with respect to the Earth's equator ($i$) in blue, and the obliquity of the Earth with respect to
the ecliptic ($\varepsilon_p$) in green. In the right panels we plotted the evolution of the Moon's nodes on the ecliptic ($\Omega$) in
red, and the node of the Earth's equator on the ecliptic ($\psi$) in green. We distinguish three types of motion, as outlined in Atobe
\& Ida (2007). On the top panels, the Moon's orbit and the Earth's spin precess around $\bf{k}_p$, which results in $I$ and
$\varepsilon$ being nearly constant, and $i$ oscillating between $\varepsilon_p - I$ and $\varepsilon_p+I$ with a period equal to the
revolution of $\Omega$ ($\sim$ 18.5~yr). On the bottom panels, the Moon's orbit precesses about the common angular momentum vector of
the spin of the Earth and the lunar orbit, $\bf{s}_p + \bf{k}_s$. Thus, the mutual inclination, $i$, stays almost constant while $I$
and $\varepsilon_p$ oscillate out of phase with a period equal to half the precession period of the satellite's orbit on the equator of
the planet (e.g. Kinoshita \& Nakai, 1991). Superimposed on this is a long-period precession of $\bf{k}_s + \bf{s}_p$ (bottom-right
panel). In the middle panels, the precession of both $\bf{s}_p$ and $\bf{k}_s$ is neither around their sum nor around $\bf{k}_p$, but
somewhere in between (Atobe \& Ida, 2007). Thus $I$, $i$ and $\varepsilon_p$ all oscillate with large amplitude. Here we find that the
precession period of the Earth's axis is about 500~yr; Bou\'{e} \& Laskar (2006) found it was 450~yr.\\

\begin{figure*}
\resizebox{\hsize}{!}{\includegraphics[angle=-90]{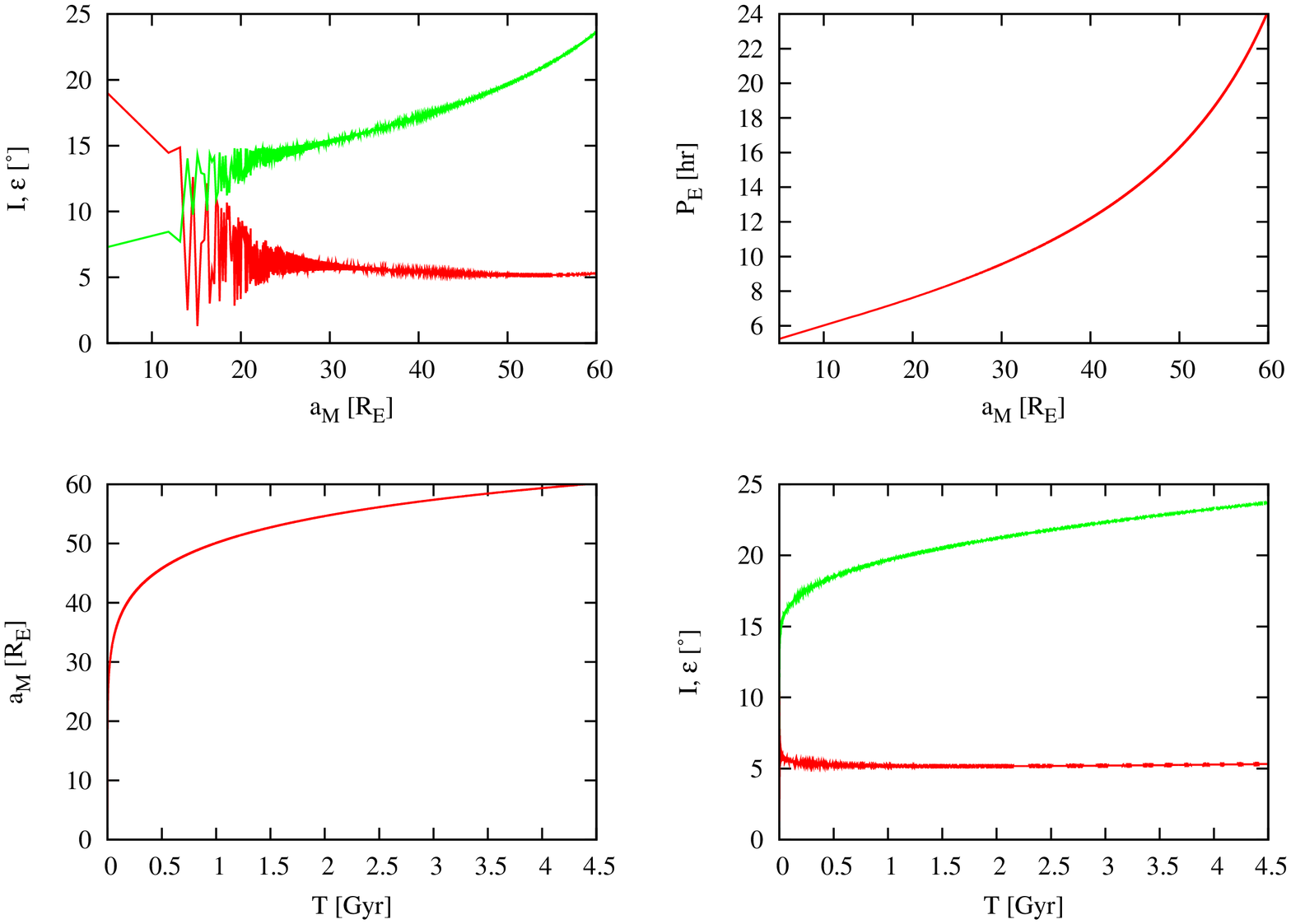}}
\caption{Tidal evolution of the Earth-Moon system. This plot should be compared with Touma \& Wisdom (1994). The top-left panel plots
$I$ (red) and $\varepsilon_p$ (green) vs $a_M$. The top-right panel shows the Earth's rotation period vs $a_M$. The bottom-left panel
shows $a_M$ vs time and the bottom-right panel depicts $I$ (red) and $\varepsilon_p$ (green) vs time.}
\label{emtides}
\end{figure*}

The tidal evolution of the Earth-Moon system to the present is shown in Fig.~\ref{emtides}, with initial $m_s = 0.01223$~$m_{\oplus}$,
$\varepsilon_p = 7.3^\circ$ and $I=19^\circ$. The values of the other relevant quantities are in Table~\ref{tab1}. The top-left panel
depicts the obliquity of the Earth (green) and the inclination of the Moon with respect to the ecliptic (red) versus the semi-major
axis of the lunar orbit in Earth radii. To reproduce the current lunar ecliptic inclination, a high original inclination close to the
Earth is needed, $i \sim 11^\circ$ (Touma \& Wisdom, 1994). This appears in contradiction with lunar formation theory from an impact
(Canup, 2004). A possible solution to this problem is if the Moon passed through the evection and eviction resonances at 4.6~$R_\oplus$
and 6~$R_\oplus$ (Touma \& Wisdom, 1998), or maybe if the Earth originally had two moons (Jutzi \& Asphaug, 2011). \\

The top-right panel of Fig.~\ref{emtides} depicts the rotation period of the Earth vs the semi-major axis of the Moon. The results of
both the top panels agree with those presented in Touma \& Wisdom (1994). The bottom-left panel depicts the semi-major axis of the
lunar orbit with time. It is well known that the current dissipation in the Earth is anomalously high (e.g. Williams, 1999), caused
by a near-resonance between ocean free modes and tidal forcing (Webb, 1982). Here we have just scaled the time such that the current
system is reproduced at 4.5~Gyr. Unlike the tidal model with constant phase lag, where $Q$ is independent of the tidal forcing
frequency, for the Mignard tidal model the semi-major axis evolution is not expressible in closed form as a function of time. At
small semi-major axis $a_s(t) \propto t^{0.15}$ and for large semi-major axis $a_s(t) \propto t^{0.12}$, similar to, but not the same
as for the constant phase lag model where $a_s(t) \propto t^{2/13} \approx t^{0.153}$. In the same figure, the bottom-right panel
depicts $I$ and $\varepsilon_p$ as functions of time: $I$ decreases monotonically to about 5$^\circ$ while $\varepsilon_p$ continues to
increase.\\

\begin{figure}
\resizebox{\hsize}{!}{\includegraphics[angle=-90]{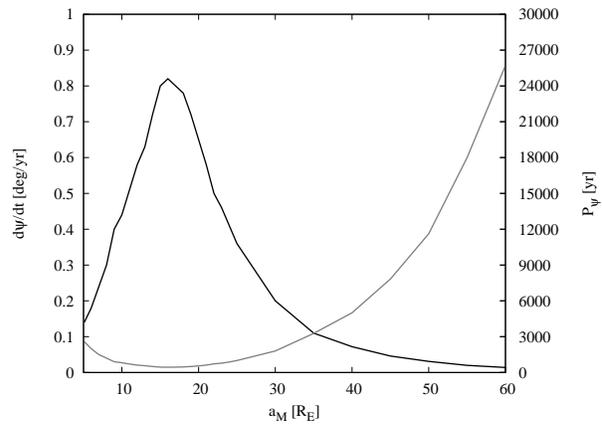}}
\caption{Precession frequency, $\dot{\psi}$ (black line) and the period of revolution of the Earth's spin axis (grey line) as a
function of lunar separation.}
\label{emprec}
\end{figure}

The result of a final test is depicted in Fig.~\ref{emprec}, which plots the precession frequency of the Earth's spin (black line)
and the corresponding precession period (grey line) as a function of lunar distance. The data points were generated from numerical
simulations: we tidally evolved the Earth-Moon system up to a pre-determined value of the semi-major axis of the Moon, using the
initial conditions of Table~\ref{tab1} with initial $\varepsilon_p = 7.3^\circ$ and $I=19^\circ$, similar to Touma \& Wisdom (1994).
Once the Moon had evolved to the designated distance, the simulation was stopped and the final state was integrated for 100~kyr without
tides. The general shape and magnitude of the curve agrees with Touma \& Wisdom (1994) and Bou\'{e} \& Laskar (2006), but the maximum
is at a larger lunar semi-major axis. We believe this is caused by us changing the value of the Earth's $J_2$ while Bou\'{e} \& Laskar
(2006) kept it constant. \\

Now that we have demonstrated the validity of our codes through comparison with earlier work, we performed a series of simulations
similar to Atobe \& Ida (2007): we fixed all the input quantities of the system to their values listed in Table~\ref{tab1}, but
changed the initial obliquity of the planet and the mass of the satellite. The obliquity was increased from 0 to 175$^\circ$ in steps
of 5$^\circ$ while the satellite mass ranged from 0.0025~$m_p$ to 0.05~$m_p$ in steps of 0.0025~$m_p$. Each of these planet-satellite
systems were integrated for 4.5~Gyr and their endstates were recorded. These endstates were subsequently integrated for 100~kyr without
tides to determine the spin precession frequency of the planet and compare them with analytical results of Bou\'{e} \& Laskar (2006).
These endstates should quantitatively reflect the possible configurations of the planet-satellite system for different values of the
initial satellite mass and initial planetary obliquity. While the outcome is similar to Atobe \& Ida (2007), our approach is different
because we terminate the simulations at a given time rather than determine the final state of the system after all tidal evolution. We
also checked whether the planet's precession frequency will pass through a secular spin-orbit resonance with one or more of the
inclination eigenfrequencies of the solar system, $s_i$ (Brouwer \& van Woerkom, 1950). Such a resonance would induce a large, sudden
increase in the planet's obliquity. We saved the state of the system whenever such a resonance was crossed for future analysis.

\section{Results}
In this section we present the results from our numerical simulations. We have used the third version of our code i.e. the one
implementing both satellite and solar tides. We show the final state of the planet-satellite system at 4.5~Gyr in the form of several
figures. The results of simulations with different initial planetary spin periods and planetary semi-major axis are shown further
below. Whenever the satellite collided with the planet the simulation was stopped and the final obliquity and spin period of the
planet were recorded. No subsequent solar tidal evolution was taken into account.\\

\subsection{Endstates}

\begin{figure}
\resizebox{\hsize}{!}{\includegraphics[angle=-90]{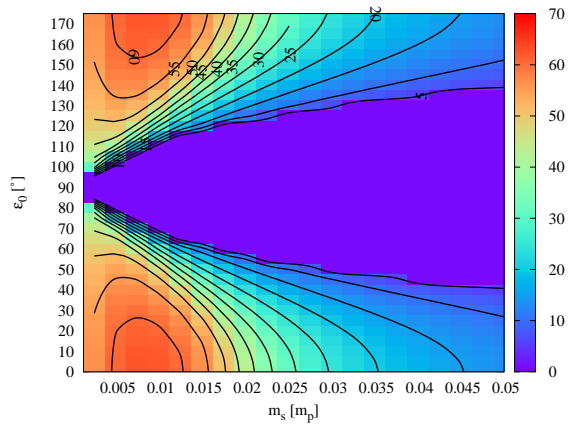}}
\caption{Contour plot of the satellite's semi-major axis (in planetary radii) after 4.5~Gyr of tidal evolution. In the central
wedge-shaped region the satellite collides with the planet shortly after its formation (Atobe \& Ida, 2007).}
\label{amoon4G}
\end{figure}

In Fig.~\ref{amoon4G} we plot the semi-major axis of the satellite in planetary radii after 4.5~Gyr of tidal evolution as a function of
the mass of the satellite and the initial obliquity of the planet. All subsequent figures will be of this type. The colour coding on
the right shows the scale. The deep purple region in the middle of the figure, which flares out with increasing satellite mass, is the
region where the satellite falls onto the planet (Atobe \& Ida, 2007). In all the figures the collision region will be this colour.
This endstate occurs for $\varepsilon_0$ ranging from approximately $60^\circ$ to $120^\circ$. Here extreme seasonal variations in
insolation and temperature may occur (e.g. Williams \& Kasting, 1997) and these could be a concern for terrestrial-type life. The
widest satellite orbits are obtained for low-obliquity planets and masses between 0.005$m_p$ and 0.0125$m_p$, and for planets with
initial obliquity close to 180$^\circ$. Smaller satellites do not raise a high enough bulge on the planet and thus need more time to
expand. It should be noted that in all figures, the structure is symmetrical around $\varepsilon_0 = 90^\circ$. The size of the
satellite orbit affects the strength of the tidal bulge on the planet and its pole precession frequency. The latter, especially, could
be important for habitability when this frequency is close to one of the eigenfrequencies of the solar system. \\

\begin{figure}
\resizebox{\hsize}{!}{\includegraphics[angle=-90]{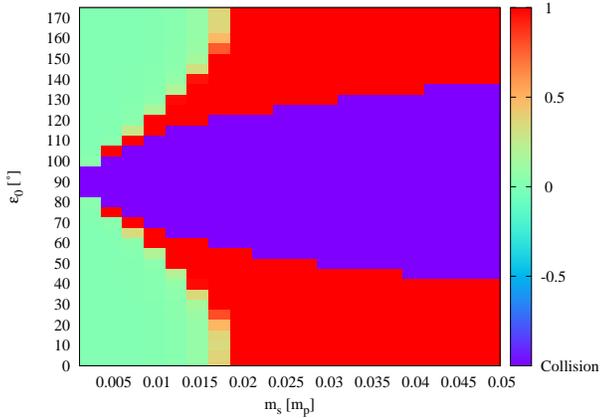}}
\caption{Colour surface plot of the period ratio of the planet's rotation and the satellite orbit. A ratio of 1 means the system is in
the double synchronous state.}
\label{sync4G}
\end{figure}

Something different occurs for massive satellites, which is clearly seen in Fig.~\ref{sync4G}. Here we plot the ratio of the rotation
period of the planet to the orbital period of the satellite. For satellites more massive than 0.02~$m_p$ the system is in the
double synchronous state. Atobe \& Ida (2007) call this evolution type C. The final period ratio depends steeply on $m_s$ and
scales as $m_s^{14}$ (Atobe \& Ida, 2007) and thus the transition from an evolving system to a double synchronous one with increasing
$m_s$ is rather abrupt. Applying this to our own system suggests that the habitability of the Earth would be completely different
if the Moon were just 50\% more massive. The double synchronous state also explains the decreasing final satellite semi-major axis with
increasing satellite mass: the system reaches the double synchronous state and the final orbit is closer to the planet because of the
lower initial ratio of angular momentum in the planet's rotation vs. that in the satellite's orbit. \\

\begin{figure}
\resizebox{\hsize}{!}{\includegraphics[angle=-90]{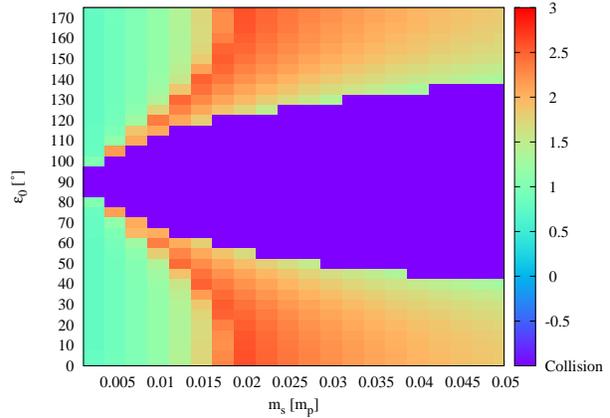}}
\caption{Colour surface plot of the logarithm of the planet's final rotation period in hours.}
\label{prot4G}
\end{figure}

The corresponding rotation period of the planet after 4.5~Gyr of tidal evolution is shown in Fig.~\ref{prot4G}. Here we plot
$\log(P_r/1\,{\rm{hr}})$ as a function of satellite mass and initial planet obliquity. Similar to Fig.~\ref{sync4G} above, there is a
very steep increase of the rotation period as a function of satellite mass, and the period reaches a maximum for a mass of
approximately 0.02$m_p$. Here the rotation period of the planet is several weeks and the satellite semi-major axis sits between 40 and
45~$R_p$. As the satellite mass increases, the final rotation period of the planet is reduced because of the larger fraction of
initial angular momentum residing in the satellite orbit compared to the planet's rotation. Even so, in the double synchronous state
the rotation period of the planet is always longer than 96~hours, apart from very close to the edge of the central wedge. \\

As we discussed earlier, these long rotation periods affect the climate on the planet and have implications for the origin and
sustenance of life. In the double synchronous case the rotation period of the planet could be a substantial fraction of the relaxation
time scale and large diurnal temperature variations could occur. Furthermore, in the double synchronous state there are no tides and
thus ocean mixing could stop, thereby reducing possible heat transport to the poles. However, the slow rotation rate may increase the
Hadley cell to reach the poles, which increases the heat flow to and the temperature at the polar regions. More studies are needed to
decide which of these two effects dominates over the other. Lastly, the long rotation period reduces the $J_2$ moment and causes
the precession period of the planet to increase. This issue is discussed below.\\

Systems with very light satellites (say $m_s < 0.005$~$m_p$) may negatively affect the habitability because the lack of a tidal
bulge raised by the satellite causes the planet to still rotate rapidly, which reduces ocean mixing. This would decrease the heat flow
towards the poles, likely causing sustained low temperatures there that reduce habitability.\\

\begin{figure}
\resizebox{\hsize}{!}{\includegraphics[angle=-90]{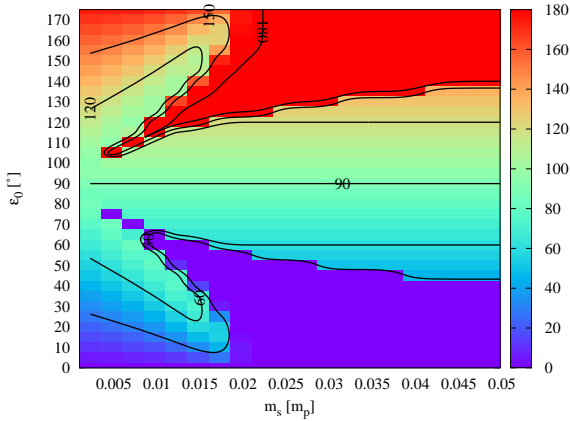}}
\caption{Contour plot of the final obliquity of the planet in degrees.}
\label{obl4G}
\end{figure}

Even though many systems are in the double synchronous state after 4.5~Gyr, the real test to measure if the system has fully evolved
is to determine the final obliquity of the planet. From Atobe \& Ida (2007) we know that when the double synchronous state is reached,
the planet's obliquity slowly decreases back to 0 if prograde, or increases to 180$^\circ$ if retrograde. Figure~\ref{obl4G} depicts
the obliquity of the planet after 4.5~Gyr of tidal evolution. All double synchronous systems are fully evolved and the planet's
obliquity is at 0 or 180$^\circ$. The only cases where the obliquity has a different value is when the system is not synchronous or
when the satellite has collided with the planet. For planets whose spin is perpendicular to their orbit, the habitable regions are
substantially reduced from planets with moderate to high obliquities (Spiegel et al., 2009). \\

\begin{figure}
\resizebox{\hsize}{!}{\includegraphics[angle=-90]{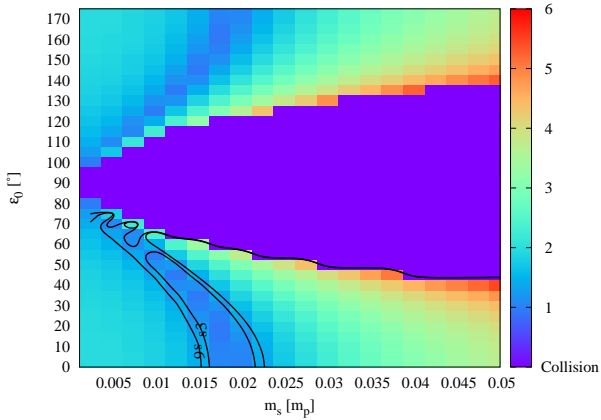}}
\caption{Contour plot of the logarithm precession frequency, in arcsec per year, of the planet's spin. Contours for resonances with the
eigenfrequencies $s_6$ and $s_3$ are indicated.}
\label{prec4G}
\end{figure}

One last quantity that determines the stability of the obliquity, and potentially the habitability, is the precession frequency of the
spin pole of the planet. Figure~\ref{prec4G} plots the logarithm of the precession frequency of the spin pole of the planet, in
arcsec per year, as a function of satellite mass and initial obliquity. In order to have a rough idea of whether this system will
experience a secular spin-orbit resonance, contours corresponding to resonances with the nodal eigenfrequencies $s_6$ and
$s_3$ are indicated. Since $s_4 \sim s_3$ (e.g. Brouwer \& van Woerkom, 1950) we did not include it here. We realise that in a
system different from our own Solar system, the eigenfrequencies should not be the same, and thus the contours serve an illustrative
purpose only. The precession of the planet is slowest when the rotation period is the longest, and some of these systems could have
evolved through the $\dot{\psi}=s_6$ and $\dot{\psi}=s_3$ and $\dot{\psi} = s_4$ secular spin orbit resonances.\\

\begin{figure}
\resizebox{\hsize}{!}{\includegraphics[angle=-90]{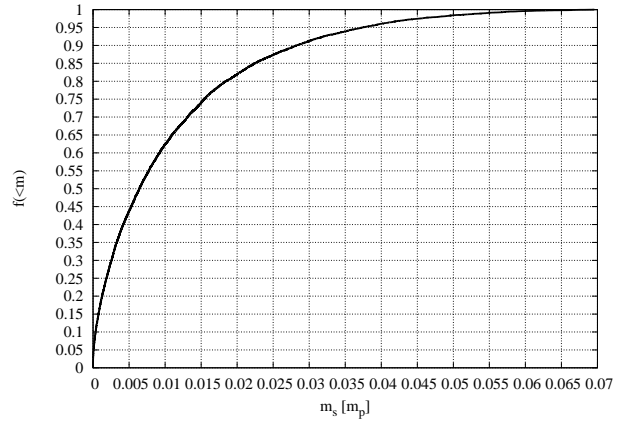}}
\caption{Cumulative distribution of satellite mass generated through a giant impact. See text and Appendix A for details.}
\label{mdist}
\end{figure}

From the above figures it appears there are generally three outcomes: i) the system is still evolving, ii) the system is in the
double synchronous state, or iii) the planet has no satellite (initial planetary obliquity near 90$^{\circ}$). The current Earth-Moon
system belongs to the first category. We can estimate the probability of being in the first state once we know the mass distribution of
the satellites from giant impacts. We have performed this analysis and detailed it in Appendix A and displayed the result in
Fig.~\ref{mdist}. The probability for the system to be still evolving is the probability of the system to not be synchronous, and
can be evaluated as

\begin{equation}
 P_H = \frac{\sum_H p_{\varepsilon_0}p_m}{\sum_T p_{\varepsilon_0}p_m}
\end{equation}
where $p_{\varepsilon_0}$ is the probability of the initial obliquity to be $\varepsilon_0$ and $p_m$ is the probability of the
satellite to have a mass $m_s$. The numerator sums over all the non-synchronous cases while the bottom sum is for all cases. The
probability of the initial spin has the distribution $p_{\varepsilon_0} d\varepsilon_0 = \frac{1}{2}\sin \varepsilon_0
d\varepsilon_0$ (Kokubo \& Ida, 2007). We find that the total probability that the system is still evolving is 85\%.\\

However, what systems yield a state similar to our own i.e. they have $12 < P_r < 48$~h, $\varepsilon_p < 40^\circ$ $(>140^\circ)$ and
$0.005 < m_s < 0.02$~$m_p$? From examining Fig.~\ref{obl4G} we can find the initial obliquity that yields a system with the final
planetary obliquity and spin period in the specified ranges. We can then repeat the same procedure as above, which yields a probability
of 14\% for the system to still be evolving, have $\varepsilon_p<40^\circ$ or $\varepsilon_p>140^\circ$ and $12<P_r<48$~h. This
value is somewhat uncertain because there is no unique method to obtain this probability, and thus it only serves as a rough
estimate.\\

There are several effects that we have not taken into account, such as the planet's distance from the Sun, the initial rotation period
of the planet and the effect of the perturbations of the other planets on the obliquity of the planet. Each of these will be discussed
in turn.

\subsection{Secular spin-orbit resonance crossing}
It is unlikely that extrasolar terrestrial planets will exist in isolation, and recent data confirms this hypothesis (Lissauer et
al., 2012). The existence of multiple planets in a given system raises the possibility of the system crossing secular-spin orbit
resonances. The chances of that happening depends on the secular architecture of the system. Atobe et al. (2004) studied the effect of
the obliquity evolution of terrestrial exoplanets that were perturbed by a giant planet. They concluded that the terrestrial planet's
obliquity variations were too large to sustain life if the giant planet was closer than about 5 Hill radii from the terrestrial planet.
However, their study was necessarily limited to a few representative cases of planetary configurations because a general study is
currently unfeasible. \\

The only planetary system for which we know the secular eigenfrequencies with a decent precision is our own. Ideally we would like
to test how a planet-satellite system responds to a secular spin-orbit resonance crossing in a multitude of systems, preferably even in
a generalised manner, but that is not possible at this stage. Thus in what follows we shall focus on a fictitious planet-satellite
system crossing a secular spin-orbit resonance in our own solar system as a proof of concept, and cautiously use it as a possible
outcome for other systems. If the mutual inclinations of the planets in other systems is small, the outcome presented here should be
quantitatively similar to what happens in other systems.\\

Figure~\ref{prec4G} showed that it is possible for the planet-satellite system to cross several secular spin-orbit resonances, mostly
$\dot{\psi}=s_6$ and $\dot{\psi}=s_3$. The effect of such a resonance crossing is to increase the planet's obliquity. During a secular
spin-orbit resonance the obliquity oscillates around its resonant value, $\varepsilon_r$. While in resonance, the maximum libration
amplitude of the obliquity is given by (Atobe et al., 2004)

\begin{equation}
 |\Delta \cos \varepsilon_r|_{\rm{max}} = \sqrt{2N_i\sin 2\varepsilon_r},
\end{equation}
where $N_i$ is the forced inclination on the planet's orbit corresponding to the frequency $s_i$. If the oscillation range is small,
then we can use the approximation $|\Delta \varepsilon_r|_{\rm{max}} = 2\sqrt{N_i\cot \varepsilon_r}$. The increase in obliquity when
crossing a resonance is then $\sim 2|\Delta \varepsilon_r|$. Typically $N_i\cot \varepsilon_i = O(10^{-3})$ to $O(10^{-2})$ depending
on the resonant obliquity and the forcing, and thus typically $\Delta \varepsilon_r \sim 5^\circ$-10$^\circ$.\\

\begin{figure}
\resizebox{\hsize}{!}{\includegraphics[angle=-90]{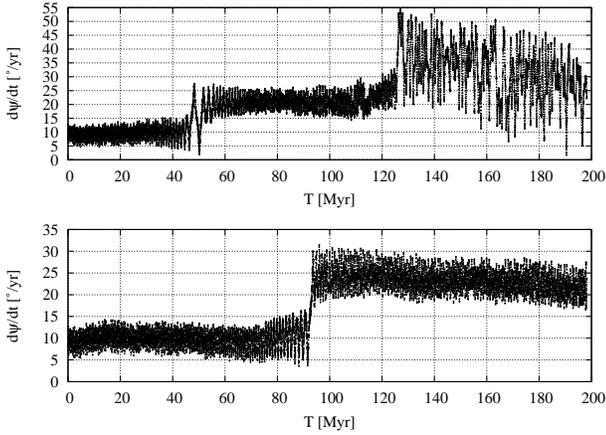}}
\caption{Obliquity evolution of the planet when passing through the $\dot{\psi}=s_6$ resonance for two different expansion rates of the
satellite orbits. The jump in obliquity is consistent with the theory presented in Atobe et al. (2004). The large obliquity ranges near
the right at the top panel are caused by the obliquity crossing the resonances $\dot{\psi}=s_3$ and $\dot{\psi}=s_4$. These resonances
overlap and the motion is chaotic.}
\label{rescross}
\end{figure}
In Fig.~\ref{rescross} we portray the obliquity of the planet vs time as it goes through the resonance crossing $\dot{\psi}=s_6$. The
precession and obliquity of the planet were integrated using the method described in Brasser \& Walsh (2011), and the precession
constant was artificially enhanced so as to match the precession rate of the planet just before it crosses the resonance with $s_6$.
During the integration the precession constant was linearly decreased in order to mimic the regression of the satellite from the
planet. In the top panel the regression was twice as fast as in the bottom panel. One can see that, when crossing the resonance, the
obliquity jumps by about 10$^\circ$ in a few million years in the bottom panel while it completes one resonant oscillation in the top
panel. In the rightmost part of the top panel the planet hits the resonance $\dot{\psi}=s_3$ and the obliquity oscillates chaotically
with large amplitude. For a low-obliquity planet the crossing through the resonance with $s_6$ could have disastrous consequences if it
has polar ice caps such as the Earth because the increased insolation at high latitudes would partially melt these ice caps, raise sea
levels and potentially wipe out a substantial portion of coastal life. Thus we argue that for terrestrial-type life to develop and be
sustained, passage through secular spin-orbit resonances should be avoided.\\

At present the precession rate of the Earth's spin pole is $\dot{\psi} = -50.5$~$''$~yr$^{-1}$. However, a 50\% increase in the lunar
mass or initial obliquity of the Earth would have resulted in the Earth-Moon system having passed through the $\dot{\psi} = s_6$
resonance before 4.5~Gyr.

\subsection{The effect of initial planetary rotation period}
\begin{figure}
\resizebox{\hsize}{!}{\includegraphics[angle=-90]{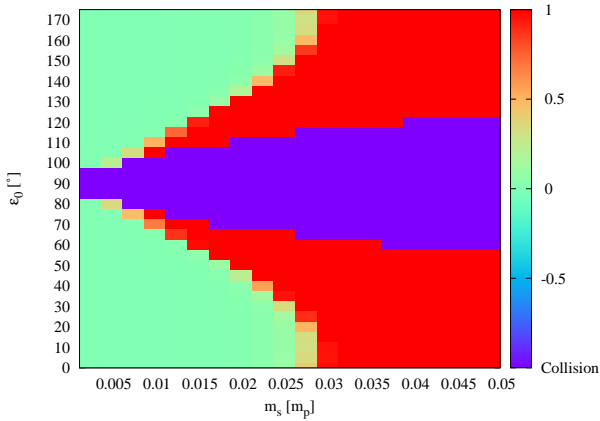}}
\caption{Colour surface plot of the period ratio of the planet's rotation and the satellite orbit. The planet's initial spin period is
3~h. A ratio of 1 means the system is in the double synchronous state.}
\label{s43h}
\end{figure}

\begin{figure}
\resizebox{\hsize}{!}{\includegraphics[angle=-90]{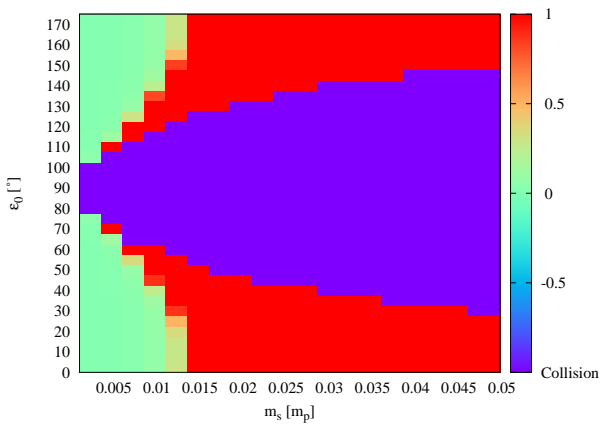}}
\caption{Colour surface plot of the period ratio of the planet's rotation and the satellite orbit. The planet's initial spin period is
7~h. A ratio of 1 means the system is in the double synchronous state.}
\label{s47h}
\end{figure}

Recent simulations of terrestrial planet formation with a realistic accretion scenario demonstrated that these planets have rotation
periods of about twice their minimum value, with $P_{\rm{min}}=2\pi\sqrt{R_p^3/Gm_p}$ and a spread of about $P_{\rm{min}}$ (Kokubo \&
Genda, 2010). In order to account for the different initial rotation period of the planet we performed a series of simulations where we
set the planet's initial period equal to 3~h and 7~h. Rather than show the full results we decided to only show the regions where the
system is doubly synchronous. The results are plotted in Figs~\ref{s43h} and~\ref{s47h}. In the case the initial rotation period is 3~h
the minimum satellite mass for which the system is synchronous has now increased beyond 0.03~$m_p$, while it is close to 0.0125~$m_p$
for the case where the initial rotation period was 7~h. Another interesting feature is that the area in the plot where the satellite
falls onto the planet also depends on the initial rotation period, and it grows as the initial period increases. The reason for this
behaviour is that the tidal evolution increases the obliquity of the planet as the satellite recedes from the planet. At high
obliquity the satellite will eventually turn around and fall onto the planet, but this only happens if the inclination between the
spin of the planet and the orbit of the satellite $i=\arccos({\bf s}_p\cdot{\bf k}_s)>90^\circ$ part of the time. The planet's
obliquity increase is lower for fast-spinning planets, a smaller number of cases will experience retrograde motion and the
satellite will not fall onto the planet.

\subsection{The effect of initial planetary semi-major axis}
\begin{figure}
\resizebox{\hsize}{!}{\includegraphics[angle=-90]{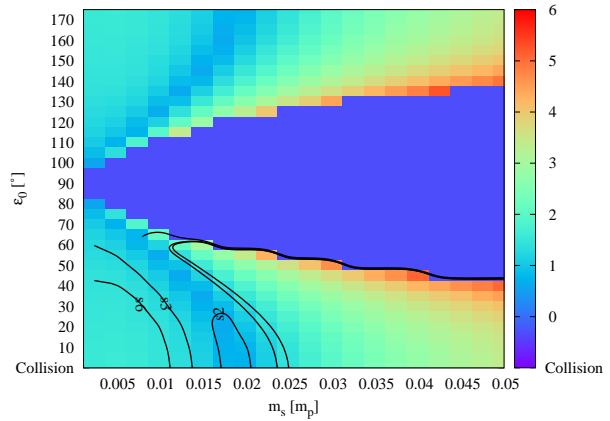}}
\caption{Contour plot of the logarithm precession frequency, in arcsec per year, of the planet's spin for systems with semi-major axis
1.25~AU. Contours for resonances with the eigenfrequencies $s_6$, $s_3$ and $s_2$ are indicated.}
\label{4g125}
\end{figure}

The habitable zone around the Sun is thought to reside between 0.7~AU and 1.3~AU (Williams \& Kasting, 1997). The precession time
of the planet's spin and the satellite's orbit scales with the planet's semi-major axis as $a_p^{-3}$. This strong dependence means
that at the edges of the habitable zone the strength of the solar tides varies by up to a factor of 3 from the value at 1~AU. Thus the
regression rate of the planet's spin and the satellite's nodes vary by the same amount, regressing twice as fast at 0.7~AU compared
to 1~AU and twice as slow at 1.25~AU compared to 1~AU. Since the final state of the system is dominated by the tidal interaction
between the planet and the satellite rather than by the influence of the Sun, we found that the final outcome is very
similar to what was presented in the figures above. However, the different precession rates of the planet's spin and the satellite's
nodes are noteworthy, in particular in the case where the planet is farther from the Sun because the region inside the $s_6$ contour of
Fig.~\ref{4g125} is greatly enhanced. For planets very close to the Sun ($a<0.5$~AU) the tidal forces from the Sun may begin to
dominate over those of the satellite for lunar mass satellites. However, at these close solar distances the Hill sphere of the planet
becomes comparable to the maximum semi-major axis of the satellite for planets with low obliquity and approximately lunar mass
satellites, and other dynamical effects cannot be ignored. Thus the investigation of these cases require a proper N-body treatment.
This is beyond the scope of the current study.\\

We have run a simulation of the tidal evolution of a planet-satellite pair where the planet has a semi-major axis 1.25~AU. Most of
the results are the same as in the previous examples, apart from the precession frequency of the planet's spin pole. Since $\dot{\psi}
\propto a^{-3}$, the precession frequency for an identical system at 1~AU should be a factor $\sim 2$ lower. This is depicted in
Fig.~\ref{4g125} above, where we show the precession frequency of the planet's spin after 4.5~Gyr of evolution as a function of the
satellite mass and initial obliquity. This should be compared directly with Fig.~\ref{prec4G}. As one may see the contours for $s_6$
and $s_3$ occupy a larger area of phase space and in some extreme cases resonance passage with $s_2$ is possible. As we have
demonstrated above, resonance with $s_3$ and $s_4$ causes large-amplitude, long-period oscillations in the obliquity and these have
drastic consequences for the climate. Thus these resonances should be avoided. \\

\section{Conclusions}
In the previous section we have presented the results from our simplified numerical simulations within a well-defined framework. Within
this framework the results are robust. We have attempted to place these results in the context of habitability of tidally-evolved
terrestrial planet-satellite systems, although the discussion is somewhat speculative due to the large uncertainties in the habitable
conditions.\\

We have performed a large sample of numerical simulations of the tidal evolution of an Earth-like planet with a satellite. The
satellite's mass and the obliquity of the planet are considered as the two free parameters; the remaining ones are modelled after the
current Earth and Moon. In our simplified model, taken from Goldreich (1966), Touma \& Wisdom (1994), Atobe \& Ida (2007) and Correia
(2009) the satellite's orbit remains circular. Rather than attempt to determine the final end state of the tidal evolution we ended
the simulations when the system reached an age of 4.5~Gyr. This age was determined by requiring that the current Earth-Moon system is
reproduced from the initial conditions of Atobe \& Ida (2007). We determined i) which systems are still evolving, ii) which ones are in
the double synchronous state accompanied by a perpendicular spin, and iii) which systems have lost their satellites. Systems
with obliquities $60^\circ \lesssim \varepsilon_p  \lesssim 120^\circ$ lose their satellites (Atobe \& Ida, 2007). We find that after
4.5~Gyr only 85\% of cases, weighted by the satellite mass derived in Appendix A, are still evolving; the rest have either lost
their satellite or have reached the double synchronous state. We also discussed habitability as a function of the planet's obliquity
and rotation period in each end state, in terms of diurnal/seasonal temperature and ocean mixing and suggest that states ii) and iii)
may be less habitable than i). Modelling accretion of a satellite from debris formed by a giant impact, we estimated the probability
for an Earth-mass planet to have the end state similar to our Earth-Moon system (12~h $< P_p <$ 48~h and $\varepsilon_0 < 40^\circ$ or
$\varepsilon_0 > 140^\circ$), which might be favourable for habitability, amounts to be only 14\%. Elser et al. (2011) conclude
that the probability of a terrestrial planet ending up with a heavy satellite ranges from 2\% to 25\% with an average of 13\%.
Combining these results suggests that the probability of ending up with a system such as our own is on average only 2\%.

\section{Acknowledgements}
The authors are deeply grateful to an anonymous reviewer who offered constructive criticism that greatly improved the quality of the
paper. Part of this work was performed while RB visited SI and EK at their respective institutes and he is grateful for their
hospitality.

\section{Appendix A: Satellite mass distribution from a giant impact}
We predict the distribution of mass of a planetary satellite accreted from an impact-generated disc. The disc is the result of a
collision between a protoplanet and a planetary embryo and the satellite will form from this disc. The results below are based on SPH
simulations of giant impacts (Canup et al. 2001; Canup 2004) and N-body simulations of accretion of satellites (Ida et al. 1997; Kokubo
et al. 2000). The purpose here is to get a rough distribution but not to pursue detailed fitting with the simulations.\\

Ida et al.(1997) and Kokubo et al.(2000) considered conservation of mass and angular momentum during the accretion:

\begin{eqnarray}
m_d & \simeq  m_{\rm{acc}}+m_{s} &    \\
L_{d} & \simeq  L_{\rm{acc}}+L_{s} & 
             \simeq m_{\rm{acc}} \sqrt{Gm_{p}R_{p}}
                  + m_{s} \sqrt{Gm_{p}a_{s}}, 
\end{eqnarray}
where $m_{d}$ and $L_{d}$ are mass and orbital angular momentum of the disc, $m_{\rm{acc}}$ and $L_{\rm{acc}}$ are those of the disc
materials that are eventually accreted to the host planet, $m_{s}$ and $L_{s}$ are those that are finally incorporated into the
satellite, $a_{s}$ is its semi-major axis, and $m_{p}$ and $R_{p}$ are a mass and a physical radius of the planet after the giant
impact. Here we neglected the disc materials that escape from the system. From these equations, 
\begin{equation}
\frac{L_{d}}{m_{d}} \simeq \left( 1 - \frac{m_{s}}{m_{d}} \right)
        \sqrt{Gm_{p}R_{p}} + \frac{m_{s}}{M_{d}} \sqrt{Gm_{p}a_{s}}.
\label{eq:disk_L}
\end{equation}

Canup et al. (2001) compiled the data of previous SPH simulations of giant impacts with $\gamma \equiv m_2/(m_2+m_1) = 0.3$ and ANEOS
for an equation of state, where $m_1$ and $m_2$ are target and projectile masses. They found that $m_{d}/m_{p}$ ($m_{p} \simeq
m_1+m_2$) and $L_{d}/L_{g}$ are given by a function of only $L_{i}/L_{g}$, where $L_{i}$ is impact angular momentum and $L_{g}$ is that
of a parabolic grazing impact. These are given by
\begin{equation}
\begin{array}{ll}
L_{i} & {\displaystyle = \frac{m_1m_2}{m_1 + m_2} R_{i} v_{\rm{esc}}\delta} \\

L_{g} & {\displaystyle = \frac{m_1 m_2}{m_1 + m_2} (R_1 + R_2) v_{\rm{esc}}
            = \mu \sqrt{2G(m_1+m_2)(R_1 + R_2) } } \\
            & {\displaystyle = \sqrt{2} [\gamma^{1/3}+(1-\gamma)^{1/3}]^{1/2} \mu
              \sqrt{G m_{p} R_{p}}},
\end{array}
\label{eq:L_graz}
\end{equation}
where $R_{i}$ is the impact parameter for a two-body encounter, $R_1$ and $R_2$ are physical radii of the target and the projectile
($R_1 > R_2$), $\mu = m_1 m_2/(m_1+m_2) \simeq \gamma (1-\gamma) m_{p}$, and we assumed that $m_{p} = m_1 + m_2$ and the internal
densities of the projectile and the target are the same. In addition from angular momentum conservation we have $\delta =
\sqrt{1+v_{\infty}^2/v_{\rm{esc}}^2}$ with $v_{\infty}$ being the speed of the impactor at the planet's Hill radius. \\

Using the compiled data in Canup et al. (2001), we found that the specific angular momentum of the disc is given approximately by
\begin{equation}
\frac{L_{d}}{m_{d}} \simeq 1.5 \frac{L_{g}}{\mu}.
\label{eq:L_specific1}
\end{equation}
Canup (2004) performed SPH simualtions of giant impacts with $\gamma = 0.11$--0.15 with the M-ANEOS (Melosh 2000) equation of
state. The new equation of state results in relatively small $L_{d}/m_{d}$,
\begin{equation}
\frac{L_{d}}{m_{d}} \simeq 1.2 \frac{L_{g}}{\mu}.
\label{eq:L_specific2}
\end{equation}
Note that Canup et al.(2001) and Canup (2004) did not present the relations (\ref{eq:L_specific1}) and (\ref{eq:L_specific2}).
We deduced these relations from their results. Substituting eqs.~(\ref{eq:L_specific1}) and (\ref{eq:L_specific2}) into equation
(\ref{eq:disk_L}), we obtain
\begin{equation}
\left( \sqrt{\frac{a_{s}}{R_{p}}} - 1 \right) m_{s}
= \left( \sqrt{2} [\gamma^{1/3}+(1-\gamma)^{1/3}]^{1/2} \alpha - 1 \right) m_{d},
\label{eq:Ms_Md}
\end{equation}
where $\alpha \simeq 1.5$ for the old ANEOS and $\alpha \simeq 1.2$ for the new ANEOS. The effects of difference in the equation of
state are expressed by a slight difference in $\alpha$.  \\

From the relations (\ref{eq:L_specific1}) and (\ref{eq:L_specific2}) it appears that an impact-generated disc would be formed from
materials of a projectile that do not overlap the target in the line of relative motion. We here follow the conventional
low-velocity oblique impact scenario (e.g., Canup et al. 2001) in order to make the discussion simple, although an impact with higher
velocity and a steeper angle could also contribute to formation of a large satellite (Reufer et al. 2012). This hypothesis allows us
to estimate $m_{d}$ from simple geometrical arguments. Here $\Delta R = R_{imp} - (R_1 - R_2)$ expresses the scale of a part of the
projectile forming the disc and $m_{d}$ can be estimated as
\begin{eqnarray}
m_{d} &\sim& \rho \left( \frac{\Delta R}{2} \right)^3
         = \frac{1}{8} \rho R_{p}^3 \left( \frac{\Delta R}{R_1+R_2} \right)^3
             \left( \frac{R_1+R_2}{R_{p}} \right)^3 \nonumber \\
         &\sim& \frac{1}{8} m_{p} \left( \frac{\Delta R}{R_1+R_2} \right)^3.
\label{eq:M_d_estimate0}
\end{eqnarray}
where we used that $[(R_1+R_2)/R_{p}]^3$ is almost constant for $\gamma \sim 0.1$--0.3. We set
\begin{equation}
m_{d} = \frac{\beta}{8} m_{p} \left( \frac{\Delta R}{R_1+R_2} \right)^3.
\label{eq:M_d_estimate}
\end{equation}
where $\beta$ is a constant of $O(1)$. From the definition of $\Delta R$,
\begin{equation}
\frac{\Delta R}{R_1+R_2} = \frac{R_{imp}}{R_1+R_2} - \frac{R_1-R_2}{R_1+R_2}
         = \frac{L_{i}}{L_{g}\delta} - \xi,
\label{eq:DelR_estimate}
\end{equation}
where 
\begin{equation}
\xi = \frac{(1-\gamma)^{1/3} - \gamma^{1/3}}{(1-\gamma)^{1/3} + \gamma^{1/3}}.
\label{eq:xi}
\end{equation}
which is $\simeq 0.14, 0.28, 0.30$ for $\gamma = 0.3, 0.15, 0.1$, respectively. From eqs.~(\ref{eq:M_d_estimate}) and
(\ref{eq:DelR_estimate}) we have
\begin{equation}
\frac{m_{d}}{m_{p}} \simeq \frac{\beta}{8} 
\left( \frac{L_{i}}{L_{g}\delta} - \xi \right)^3.
\label{eq:Md_Lgraz}
\end{equation}
If we adopt $\beta \sim 1.2$, this equation reproduces all the results in Canup et al. (2001) ($\gamma = 0.3$, the old ANEOS), 
Canup \& Asphaug (2001) ($\gamma = 0.1$, Tillotson), and Canup (2004) with ($\gamma = 0.11$-0.15). That is, the scaling relationship in
eq.~(\ref{eq:Md_Lgraz}) with projectile mass to total mass ratio ($\gamma$) and a (normalised) impact parameter $L_{i}/L_{g}$ is
successful.\\

Now we substitute eq.~(\ref{eq:Md_Lgraz}) into eq.~(\ref{eq:Ms_Md}) to obtain 
\begin{equation}
\frac{m_{s}}{m_{p}} \sim \frac{\beta}{8} \frac{\sqrt{2} [\gamma^{1/3}+(1-\gamma)^{1/3}]^{1/2} \alpha - 1}
{\sqrt{a_{s}/R_{p}} - 1} \left( \frac{L_{i}}{L_{g}} - \xi \right)^3.
\label{eq:final}
\end{equation}
Since $[\gamma^{1/3}+(1-\gamma)^{1/3}]^{1/2} \sim 1.2$--1.25 for $\gamma = 0.1$--0.3 and Ida et al.(1997) shows that $a_{s} \sim 3.8
R_{p}$, eq.~(\ref{eq:final}) reads as
\begin{equation}
\frac{m_{s}}{m_{p}} \sim C \left( \frac{L_{i}}{L_{g}\delta} - \xi \right)^3.
\label{eq:final2}
\end{equation}
with $C \sim 0.17$--0.25 ($\alpha \sim 1.2$--1.5). If one were to set $L_i=(2/5)m_p R_p^2\nu$, where $\nu$ is the planet's
rotation rate (see Section 2), together with the first equation in (\ref{eq:L_graz}) one has a relation between the mass of the
satellite and the rotation rate of the planet, though it is not a one-on-one relation. For example, we consider the case of $\gamma =
0.13$ ($\xi=0.31$) and $m_{p}=1\,m_\oplus$. In this case, a collision with $L_{i}=1.2L_{EM}$ corresponds to $L_{i}/L_{g}=0.72$
(eq.~\ref{eq:L_graz}). Then, eq.~(\ref{eq:final2}) yields $m_{s} \sim 0.014 m_\oplus \sim 1.1 m_{\rm{Moon}}$, where we used $C \sim
0.2$.\\

We used equation~(\ref{eq:final}) in a Monte-Carlo method to determine the cumulative distribution of satellite masses. We randomly
selected $\gamma \in (0.05, 0.5)$, $v_{\infty}$ was taken from a Maxwellian with maximum velocity equal to the planet's escape
velocity, and $R_i$ was chosen from $R_i \in (0.2, 1)$ in units of $R_p$ with $R_i^2$ being uniform. The resulting distribution
is shown in Fig.~\ref{mdist}.

\section{References}
Atobe K., Ida S., Ito T., 2004, Icar, 168, 223\\
Atobe K., Ida S., 2007, Icar, 188, 1\\
Barron E.~J., 1983, ESRv, 19, 305\\
Barron E.~J., 1989, Geomorphology 2, 99\\
Bice, K.,  Marotzke, J., 2002, Paleoceanography 17, 1018\\ 
Bills B.~G., Ray R.~D., 1999, GeoRL, 26, 3045\\
Bou{\'e} G., Laskar J., 2006, Icar, 185, 312\\
Brasser R., Walsh K.~J., 2011, Icar, 213, 423\\
Brouwer, D., van Woerkom, A. J. J. 1950. Astron. Papers Amer. Ephem. 13, 81-107.\\
Bulirsch, R., Stoer, J. 1966. \ Numerische Mathematik 8, 1-13.\\
Cameron A.~G.~W., 1997, Icar, 126, 126\\
Canup R.~M., Asphaug E., 2001, Natur, 412, 708\\
Canup R.~M., Ward W.~R., Cameron A.~G.~W., 2001, Icar, 150, 288\\
Canup R.~M., 2004, ARA\&A, 42, 441\\
Chambers J.~E., 2001, Icar, 152, 205\\
Correia A.~C.~M., 2009, ApJ, 704, L1\\
Dauphas, N.,  Pourmand, A., Nature 473, 489\\
Deser, C., Walsh, J.~E., Timlin, M.~S., 2000, J. Climate 13, 617\\
Egbert G.~D., Ray R.~D., 2000, Natur, 405, 775\\
Elser S., Moore B., Stadel J., Morishima R., 2011, Icar, 214, 357\\
Farrell, B. F., 1990, J. Atmos. Sci., 47, 2986\\
Garrett, C., 2003, Nature 422, 477\\
Goldreich P., 1966, RvGSP, 4, 411\\
Hartmann W.~K., Davis D.~R., 1975, Icar, 24, 504\\
Huybers, P, Wunsch, C., 2005, Nature 434, 491\\
Huybers, P., 2011, Nature 480, 229\\
Ida S., Canup R.~M., Stewart G.~R., 1997, Natur, 389, 353\\
Imbrie J., Imbrie J.~Z., 1980, Sci, 207, 943\\
Jutzi, M., Asphaug, E., 2011, Nature 476, 69. \\
Katija, K., Dabiri, J. O., 2009, Nature 460, 624\\
Kinoshita H., Nakai H., 1991, CeMDA, 52, 293\\
Kokubo E., Ida S., Makino J., 2000, Icar, 148, 419\\
Kokubo E., Ida S., 2007, ApJ, 671, 2082\\
Kokubo E., Genda H., 2010, ApJ, 714, L21\\
Lissauer J.~L., 2000, ASPC, 213, 57\\
Lissauer J.~J., et al., 2012, ApJ, 750,112\\
Matsuda, Y. 2000, Planetary meteorology Univ. of Tokyo press, Tokyo, Japan\\
Melosh H.~J., 2000, LPI, 31, 1903\\
Mignard F., 1979, M\&P, 20, 301 \\
Mignard F., 1980, M\&P, 23, 185\\
Milankovi\'{c}, M., 1941. Royal Serbian Sciences 33,633\\
Paillard D., 1998, Natur, 391, 378\\
Reufer A., Meier M., Benz W., Wieler R., 2012, Icarus, in press\\
Spiegel D.~S., Menou K., Scharf C.~A., 2009, ApJ, 691, 596\\
Touma J., Wisdom J., 1994, AJ, 108, 1943\\
Touma J., Wisdom J., 1998, AJ, 115, 1653\\
Webb D.~J., 1982, GeoJI, 70, 261\\
Williams D.~M., Kasting J.~F., 1997, Icar, 129, 254\\
Williams D.~M., Pollard D., 2003, IJAsB, 2, 1\\
Williams G.~P., 1988, ClDy, 3, 45\\
\end{document}